\documentclass[pra,twocolumn,preprintnumbers,amsmath,amssymb]{revtex4}

\usepackage{amsmath}

\usepackage{graphicx}
\graphicspath{{pict/}{}}
\newcommand\pictc[5]{\begin{figure}
            \centerline{\vspace{0mm}
\includegraphics[width=#1\columnwidth,height=0.7\textheight,keepaspectratio]{#3}}
            \protect\caption{\protect\label{fig:#4} #5}\vspace{-0mm}
                    \end{figure}            }
\newcommand\pict[4][1.0]{\pictc{#1}{!tb}{#2}{#3}{#4}}
\newcommand\rpict[1]{\ref{fig:#1}}

\newcounter{Fig}

\begin{document}
\begin{sloppy}

\title{Extraordinary transmission of nanohole lattices in gold films}

\author{Alexander Minovich$^1$, Haroldo~T. Hattori$^{2,3}$, Ian McKerracher$^2$, Hark Hoe Tan$^2$,\\
Dragomir~N. Neshev$^1$, Chennupati Jagadish$^2$, and Yuri~S. Kivshar$^1$}

\address{$^1$Nonlinear Physics Center, Research School of Physical Sciences and Engineering,\\
Australian National University, Canberra ACT 0200, Australia\\
$^2$Department of Electronic Material Engineering, Research School of Physical Sciences and Engineering,\\
Australian National University, Canberra ACT 0200, Australia\\
$^3$School of Information Technology and Electrical Engineering, University of New South Wales,\\
The Australian Defence Force Academy, Canberra ACT 2600 Australia}


\begin{abstract}
We study experimentally the transmission of light through a square lattice of nanoholes perforated in a optically-thick gold film. We observe that the periodicity of the structure enhances the light transmission for specific wavelengths, and we analyze this effect theoretically by employing finite-difference time-domain numerical simulations. Furthermore, we investigate the possibilities for manipulation of the spectral transmission in quasi-periodic and chirped lattices consisting of square nanoholes with varying hole size or lattice periodicity.
\end{abstract}

\maketitle

\section{Introduction}

The recent progress in the study of light transmission through sub-wavelength apertures~\cite{Genet:2007-39:Nat} was triggered by the discovery of extraordinary transmission of light through periodic arrays of nanoholes~\cite{Ebbesen:1998-667:Nat}, where due to the excitation of surface plasmons, the amount of light transmitted through an optically-thick film with nanoholes was several times larger than that expected for the transmission of an isolated sub-wavelength hole. In simple words, surface plasmons funneled light into these sub-wavelength apertures, which would not be normally possible because of diffraction constraints. This extraordinary transmission of light appears exceptionally promising for novel photonic spectral filters or photon sorters~\cite{Laux:2008-161:NatPhot}, but the field still requires further advances in the understanding of the processes behind this enhanced transmission in order to provide novel possibilities for its manipulation.

At the same time, the extraordinary transmission of light through subwavelength apertures indicates that light could also be confined in smaller regions beyond the diffraction limit, creating a possibility for fabrication of very compact optical devices~\cite{Atwater:2007-56:SciAm}. Currently, optical devices are significantly larger than their electronic counterparts, making the reduction of size in optical devices a very welcomed outcome of the excitation of surface plasmons.

Since surface plasmons generate very high-intensity near-fields, they could also be used to enhance nonlinear effects such as nonlinear bistability~\cite{Porto:2004-081402:PRB}, plasmon-enhanced high-harmonic generation~\cite{Kim:2008-757:Nat}, and surface enhanced Raman scattering~\cite{Kneipp:2002-597:JPhys}. Surface enhanced Raman scattering is widely used in the analysis of biological and chemical samples, and the excitation of surface plasmons can lead to a considerable enhancement of the luminescence from these samples. Recently, high-intensity electric fields were also employed to manipulate nanoparticles~\cite{Righini:2008-186804:PRL}. The manipulation of nanoparticles might be useful in drug and food control, since the manipulation can be selective with respect to the characteristics of the nanoparticle.

All these interesting and novel effects that are obtained by the excitation of surface plasmons have inspired us to further investigate the extraordinary transmission of light in lattices of nanoholes. Light transmission through these nanoholes is sensitive to the refractive index of the substrate, metal properties, and structural parameters of the perforated film~\cite{Marthandam:2007-12995:OE,Garcia:2007-10028:OE}. Changing these conditions optically can lead to novel applications in all-optical switching and light manipulation.

In this paper we study, both theoretically and experimentally, the transmission properties of gold films perforated with square subwavelength apertures in different arrangements. We show how the lattice and apertures can enhance collective plasmonic resonances in the metal film. We analyze our experimental observation by employing numerical simulations based on the finite-difference time-domain (FDTD) approach. We find that our numerical simulations are in a good agreement with the experimental results and confirm the nature of the enhanced light transmission. In addition, we explore several novel ways to design the transmission through perforated metal films by introducing quasi-periodicity or aperture size chirping.

The paper is organized as follows. In Section~\ref{sec:Experimental_studies} we discuss our experimental setup and summarize our experimental results. Section~\ref{sec:Numerical_studies} is devoted to a summary of our numerical results and a brief discussion of the numerical methods we used. In Section~\ref{sec:Quasi-periodic} we introduce and study two new geometries of the nanohole lattices, created by chirping of the square lattices by changing the hole radius or the lattice spacing. Finally, Section~\ref{sec:Conclusions} concludes the paper.

\section{Experimental studies}
\label{sec:Experimental_studies}

For fabrication of our samples we use a focused ion beam (FIB) milling system to drill nanoholes in a 200\,nm thick gold film. The gold film is deposited on top of a quartz substrate using DC magnetron sputtering in an Ar ambient. Four different sample geometries are fabricated and tested. The first two geometries represent square-hole homogeneous lattices of $40\times40$ holes with spacing of $2\,\mu$m between them and a hole size of 800 and 300\,nm, respectively. The other two samples represent somewhat the more complex geometries of Thue-Morse sequence of holes and chirped-size hole lattice, respectively.

\pict[1.05]{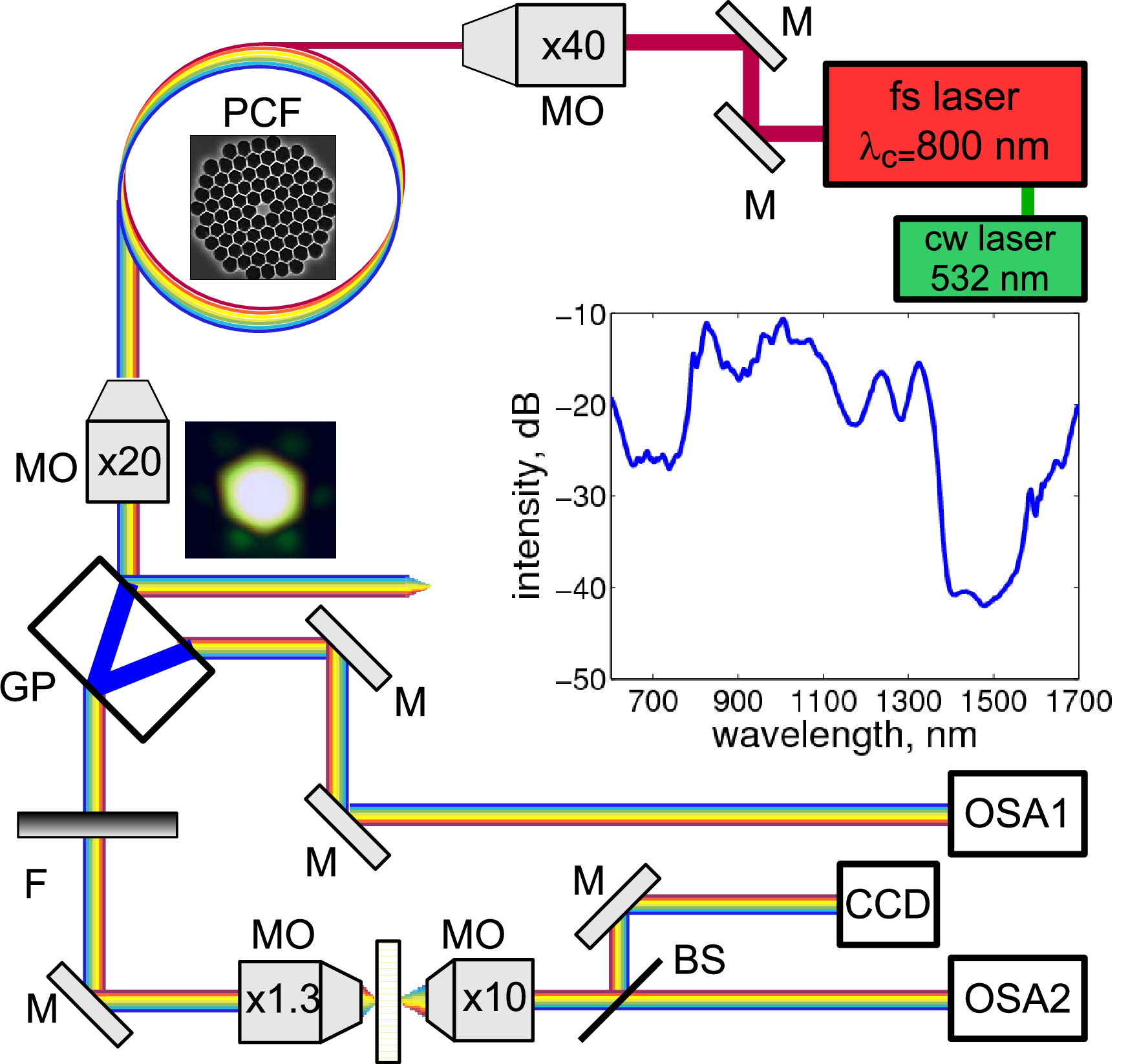}{exp_setup}{(Color online) Experimental setup: M~-- silver mirrors, MO~-- microscope objectives, PCF~-- photonic crystal fibre, GP~-- glass plate, F~--neutral density filter, BS~-- beam splitter, CCD~-- color camera, OSA~-- optical spectrum analyzers. Inset~-- graph of the generated supercontinuum spectrum.}

We characterize the spectral transmission of these samples by employing the experimental setup shown in Fig.~\rpict{exp_setup}. We use a supercontinuum radiation generated by focusing of 140\,fs pulses at 800\,nm wavelength in a photonic crystal fibre with a zero dispersion wavelength at 740\,nm. The supercontinuum spectrum spans over the wavelength range $450-1700$\,nm enabling transmission measurements over a broad spectral range. The supercontinuum beam is mildly focused onto the film from the air-gold interface.
The illuminated area of the film is 20\,$\mu$m, being determined by the size of the focal spot on the metal film. The light transmitted through the film is projected with a microscope objective on a CCD camera for identification of the perforated areas and analyzed by an optical spectrum analyser (OSA) (Agilent 86140B).
In order to minimize the influence of possible spectral fluctuations in the super-continuum source, the output spectra are recorded simultaneously with the spectrum from a reference beam on OSA\,1 (Fig.~\rpict{exp_setup}).
All measurements are normalized to the transmission through a blank quartz substrate.

\pict[0.99]{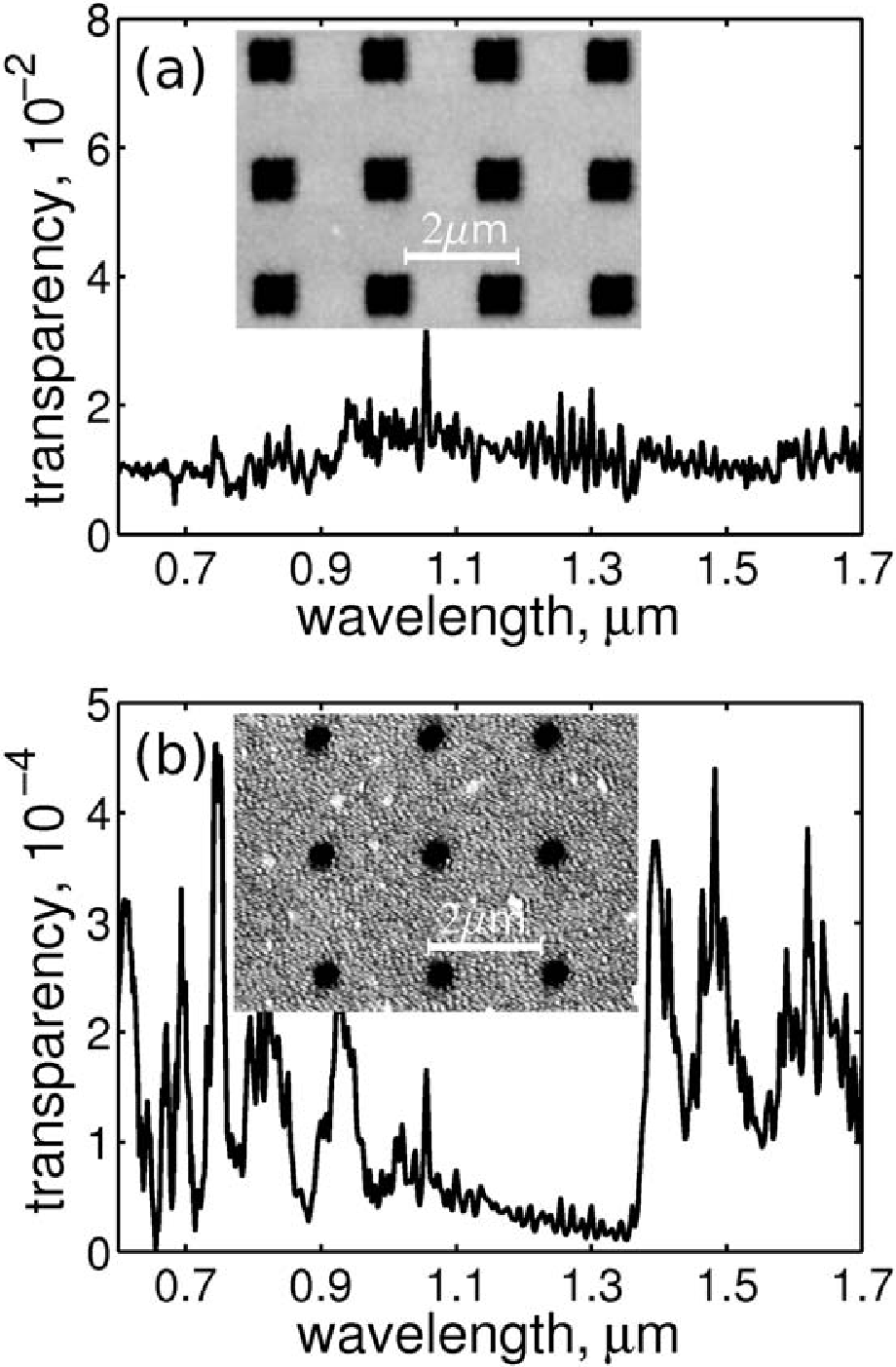}{exp_spectra}{Normalized transmission spectrum of homogenous lattice of holes with size of 800\,nm (a) and 300\,nm (b). Insets~-- atomic force microscope images of the corresponding hole lattices.}

First we measure the transmission spectra of the two homogeneous arrays consisting of 800\,nm and 300\,nm wide holes [see insets in Fig.~\rpict{exp_spectra}(a,b), respectively]. Our measurements are performed in the spectral range of $600-1700$\,nm that is determined primarily by the range of the OSA. The measured transmission spectra of both samples are shown in Fig.~\rpict{exp_spectra}. From the figure, we see that the sample with 800\,nm size holes shows, on average, higher transmission than the sample with 300\,nm size holes. However, no clear resonance enhancement peaks can be identified in the monitored wavelength range [Fig.~\rpict{exp_spectra}(a)]. On the contrary, the sample with smaller holes (300\,nm) exhibit a number of closely spaced resonances [Fig.~\rpict{exp_spectra}(b)]. Five resonances are observed in the red and near-infrared range (600\,nm--1.0\,$\mu$m) followed by monotonic drop of the transmission in the wavelength-range $1.0-1.4\,\mu$m. At even longer wavelengths (1.4--1.7\,$\mu$m) new transmission resonances can be seen. The total transmission, however, remains below $5\times10^{-4}$.

To understand the origin of the observed transmission peaks we resort to numerical calculations of the transmitted spectra. The numerical simulations also enable a direct comparison of the transmission through an array and a single aperture only, allowing to analyse the influence of periodic arrangement on the excitation of surface plasmons.

\section{Numerical studies}
\label{sec:Numerical_studies}

In our FDTD simulations, we excite the sample with a short pulse, broad spectrum source and collect the transmitted power at 3\,$\mu$m distance from the structure. At this distance, the contribution from the evanescent fields on the output transmission spectrum can be neglected. The frequency content of the transmitted power is then analyzed by utilizing fast Fourier transformation. In our simulations we use a non-uniform grid with size varying from 2\,nm near the surface to 20\,nm in a free space and a time step $cT$ of 1\,nm. To calculate the transmission through a lattice of holes we use a cell containing a single hole only and utilize periodic boundary conditions, while for transmission through a single hole we implement a perfect matching layer at the boundaries.

For the lattice of 800\,nm holes, the calculations [Fig.~\rpict{TeorSampleAB}(a)] did not reveal any clear plasmon resonances in the visible part of spectrum (apart from the pure gold transmission peak) but showed transmission peaks at infra-red wavelengths, at about 1.6, 2.3 and 3.2\,$\mu$m. This observation is in a qualitative agreement with our experiments, where no distinct resonances were measured in the transmission in the visible and near-infra-red regions.

The sample with the lattice of 300\,nm holes, on the other hand shows few weak peaks in the wavelength range between 600\,nm--1.0\,$\mu$m, followed by a lack of transmission in the region $1.0-1.3\,\mu$m, and some weaker transmission peaks in the 1.5\,$\mu$m range [Fig.~\rpict{TeorSampleAB}(b)]. These characteristic features well match the observed resonance peaks in the experimental data, however, the strength of these resonances differs from the measured quantities. Most likely this is due to imperfections in the fabricated samples, including deviation from the square shape, surface roughness [see AFM image in Fig.~\rpict{exp_spectra}(b)], or residual gold inside the holes.

\pict[0.99]{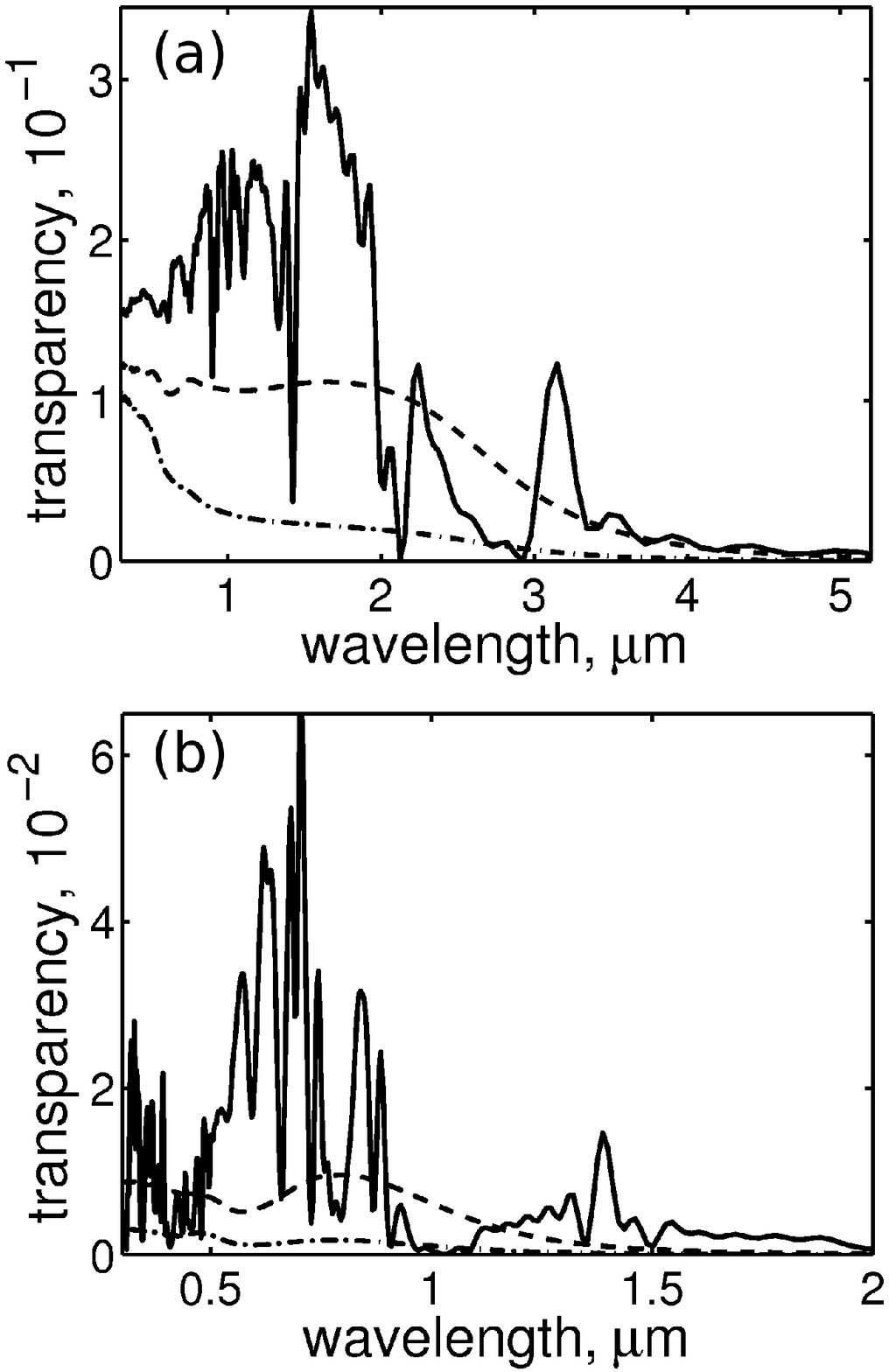}{TeorSampleAB}{Calculated transmission spectra of homogeneous square hole lattices (solid line), a single hole when light is collected by a monitor with 0.8 numerical aperture (NA) (dashed line) and with NA around 0.2 (dashed dotted line). (a)-- 800\,nm hole size; and (b)-- 300\,nm hole size.}

A comparison of the spectra for a nanoholes lattice [Fig.~\rpict{TeorSampleAB}(solid lines)] and for a single isolated hole [Fig.~\rpict{TeorSampleAB}(dashed line)] show that the observed enhanced transmission peaks are clearly due to a collective effect from the periodic arrangement of the nanoholes (Fig.~\rpict{TeorSampleAB}). In the particular, in the case of holes of 300\,nm size, we can see more than six times enhancement of the transmission in a comparison with the light transmitted through a single hole at a wavelength of 700\,nm. This enhancement is due to the excitation of surface plasmons which is allowed due to the periodic arrangement of holes. This periodicity enables the momentum matching of the incident light to the excited surface plasmons. The excitation of surface plasmons can be clearly seen in the example of electric field profiles calculated for a wavelength of 700\,nm. Figure~\rpict{SampleBModes}(a) shows the $Z$ component of the electric field after the incident light is coupled to a single square hole of 300\,nm size. One can see that the field in this case is only concentrated near the edges of the hole. In contrast, in the case of periodic arrangements of holes [Fig.~\rpict{SampleBModes}(b)], the electric field is spread in the areas between the holes, due to the excitation of surface plasmons. This excitation enables collection of light over a larger area and therefore results in enhancement of the transmission. We note that the field is higher at the air-gold interface and this feature is preserved independently if the sample is illuminated from the air or the quartz side.

\pict[0.99]{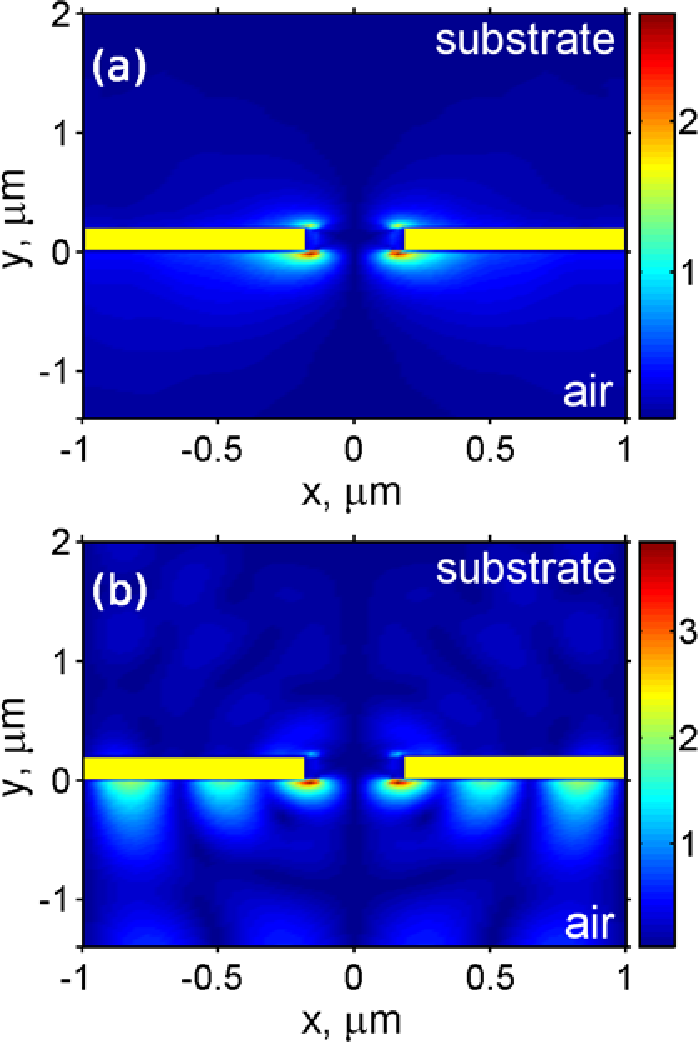}{SampleBModes}{(Colour online) Calculated amplitude distribution of $Z$ component of electrical field in simulation for a single hole (a) and periodical lattice of holes (b). The cross section is made at the center of hole. Light source is polarized along X, $\lambda=0.7\,\mu$m.}

We would also like to point out a feature of the transmission spectra which is essential for the experimental characterization and can lead to significant differences between calculations and experimental measurements. Due to the high divergence of the light exiting through the nanoholes~\cite{Schouten:2003-371:OE}, the measurement of transmission need to be realised with high numerical aperture (NA) objectives in order to collect as much light as possible from the output face. To demonstrate the importance of the output divergence of the beam, we performed calculations for a single hole transmission when we collected light by a monitor with NA of 0.8 and 0.2, as shown in Fig.~\rpict{TeorSampleAB}, dashed and dashed-dotted line, respectively. As one can see, the NA of the collecting objective can cause not only quantitative but also qualitative changes in the measured transmission spectrum.

\pict[0.99]{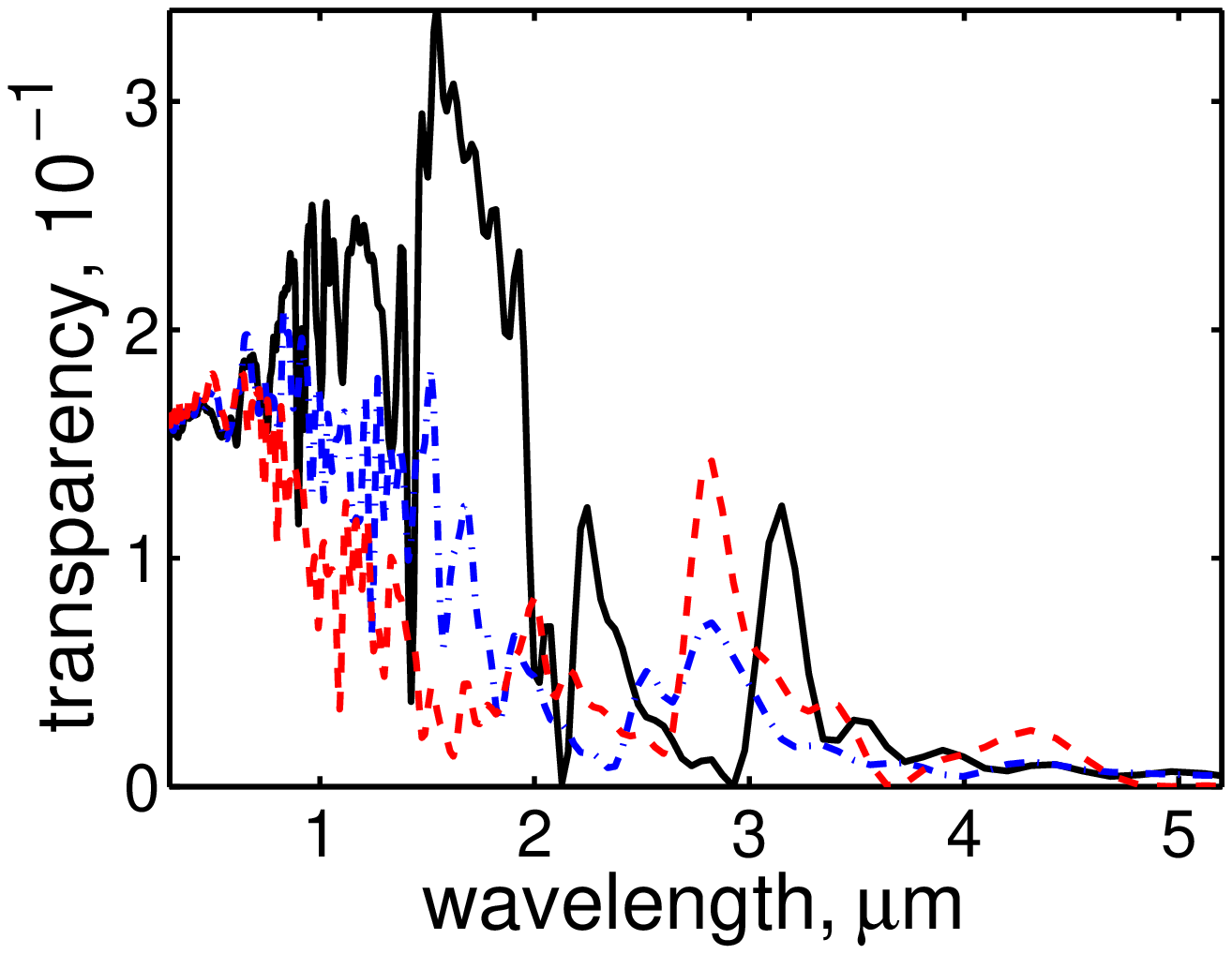}{TeorSampleAArrayAngles}{(Colour online) 800\,nm hole lattice: Transmission spectra with normal (solid line), $5^\circ$ (dashed dotted line) and $10^\circ$ (dashed line) angles of incidence.}

Next, we performed a number of simulations in order to investigate the dependence of the transmission on the angle of incidence (Fig.~\rpict{TeorSampleAArrayAngles}). In our simulations, we considered a sample with 800\,nm size square holes and calculated the spectral transmission for normal incidence and two other inclination angles. The results show that the inclination of the sample causes significant deformation of the transmission spectrum in the short-wavelength region, leading to lower transmission. Somewhat more clear is the transformation of the position of transmission peaks near 2.3 and 3.2\,$\mu$m. We see that these peaks shifts to shorter wavelength with change of angle of incidence [Fig.~\rpict{TeorSampleAArrayAngles}(dash-dotted and dashed lines) for angles of 5$^\circ$ and 10$^\circ$, respectively]. This result means that for larger transverse k-vectors of the incident light, the frequency shifts up and indicates the normal dispersion of these plasmonic modes~\cite{Ebbesen:1998-667:Nat}.

\pict[0.99]{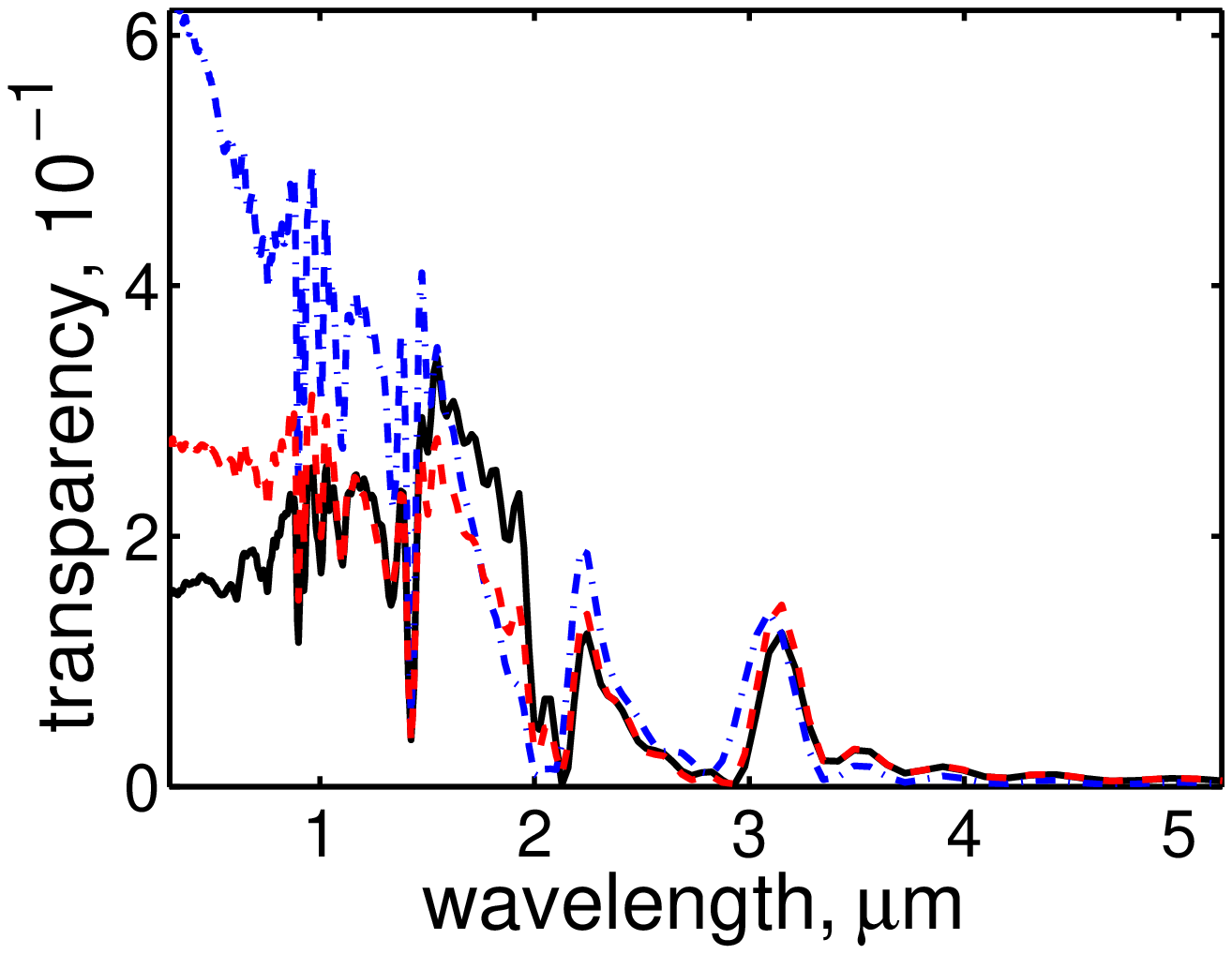}{TeorSampleAArrayGausse}{(Colour online) Transmission through a 800\,nm hole-size lattice when the film is illuminated by a plane wave (solid line), Gaussian beam waist size $1\,\mu$m (dashed dotted line) and $2\,\mu$m (dashed line). Source focal planes are located at $1.4\,\mu$m from the gold surface.}

From an experimental point of view it is also important to understand how the beam divergence would influence the transmitted spectrum. We therefore, analyzed numerically the influence of small misalignments of the focal plane near the metal surface. For this purpose, we illuminated the sample with a diverging Gaussian beam and compared the transmission measurements with those of a plane wave. In particular, we illuminate the lattice of 800\,nm holes by a Gaussian beam with a waist size of $1\,\mu$m or $2\,\mu$m and a focal plane placed at $1.4\,\mu$m from the metal surface. Our results shown in Fig.~\rpict{TeorSampleAArrayGausse} reveal that the effect of beam divergence on the spectrum is mostly at short wavelengths, where the beam size obviously increases the total transmission through the holes. In general, however, the spectral features remain the same for the diverging beams.


\pict[0.99]{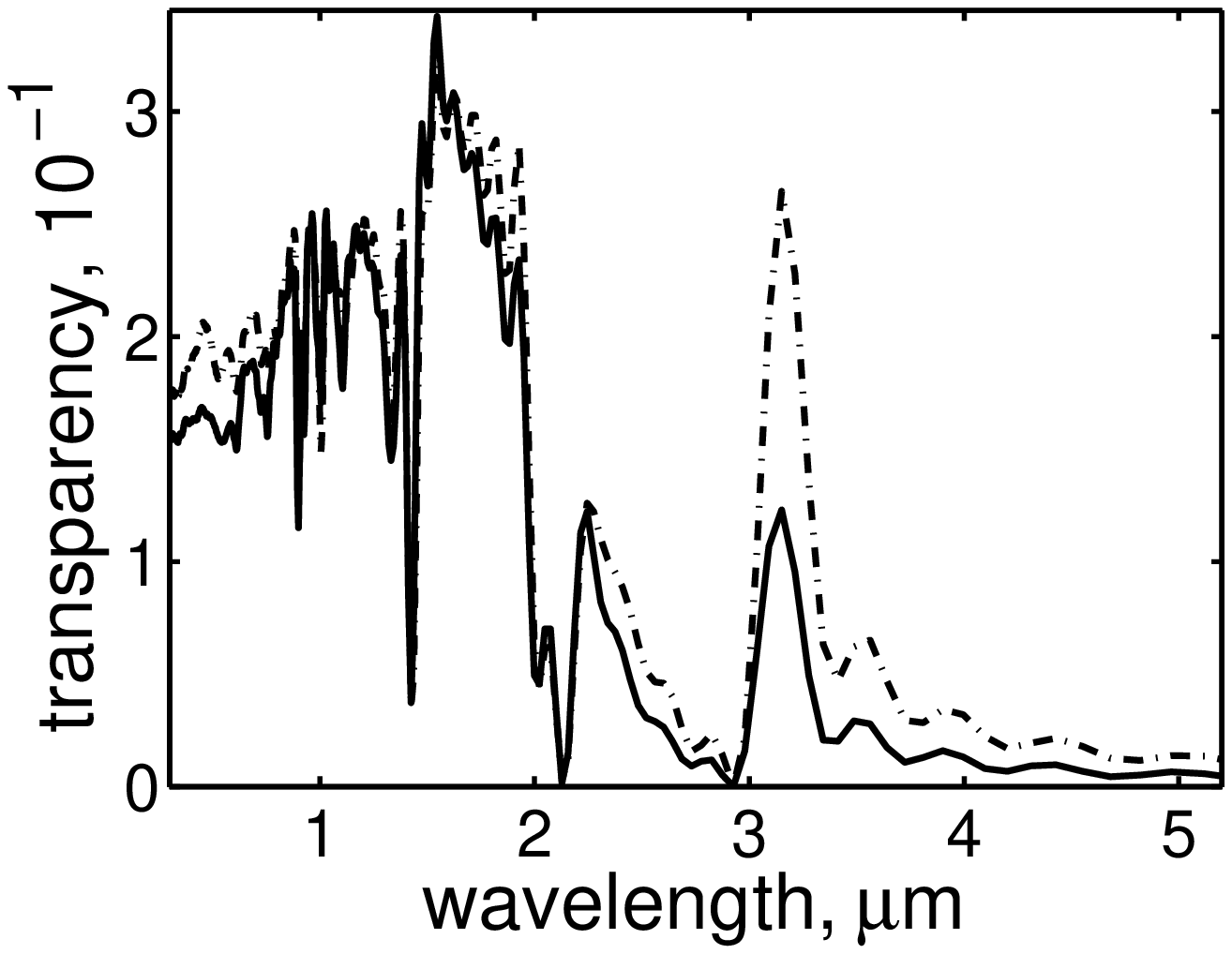}{TeorSampleAArrayThickness}{Transmission through 800\,nm hole-size lattice on a gold film of thickness 200\,nm (solid line) and 100\,nm (dash-dotted line).}

Finally, we study how the film thickness changes the transmission spectrum. In Fig.~\rpict{TeorSampleAArrayThickness} we compare the transmission spectra through the 800\,nm nanohole lattice for metal films of 200\,nm (solid line) and 100\,nm (dashed-dotted line) thickness. In agreement with previous studies~\cite{Ebbesen:1998-667:Nat}, our results show that the change of the film thickness affects only slightly the position of the peak at $3.2\,\mu$m. The influence of the film thickness is mostly expressed in reduction of the intensity of the transmitted peaks with increasing of the thickness.

\section{Quasi-periodic and chirped lattices}
\label{sec:Quasi-periodic}

As a final step, we investigate different ways to engineer the properties of the perforated films, searching for possibilities to tailor the transmission through the films. For this purpose we consider two types of nanohole structures: (i) a structure with a quasi-periodic arrangement of the holes, and (ii) a structure with constant periodicity, but with a chirped size of the holes. In the first type of structure, we preserve the transmission properties of the isolated holes, however, the quasi-periodic arrangement influences the excitation of surface plasmons. An example of such a structure is the Thue-Morse sequence pattern as shown in the inset of Fig.~\rpict{quasi-periodic}(a).

In the second type structure, we preserve the excitation of the surface plasmons (momentum matching in the transverse direction) by keeping a constant periodicity of the lattice. However, due to the hole-size chirping we significantly modify the transmission properties of the individual apertures of the lattice. An example of such a structure is shown in the inset of Fig.~\rpict{quasi-periodic}(b), which shows a lattice with the period of 2\,$\mu$m and the hole size varying in horizontal direction from 300\,nm to 800\,nm (50\,nm change from one hole to another).

\pict[0.99]{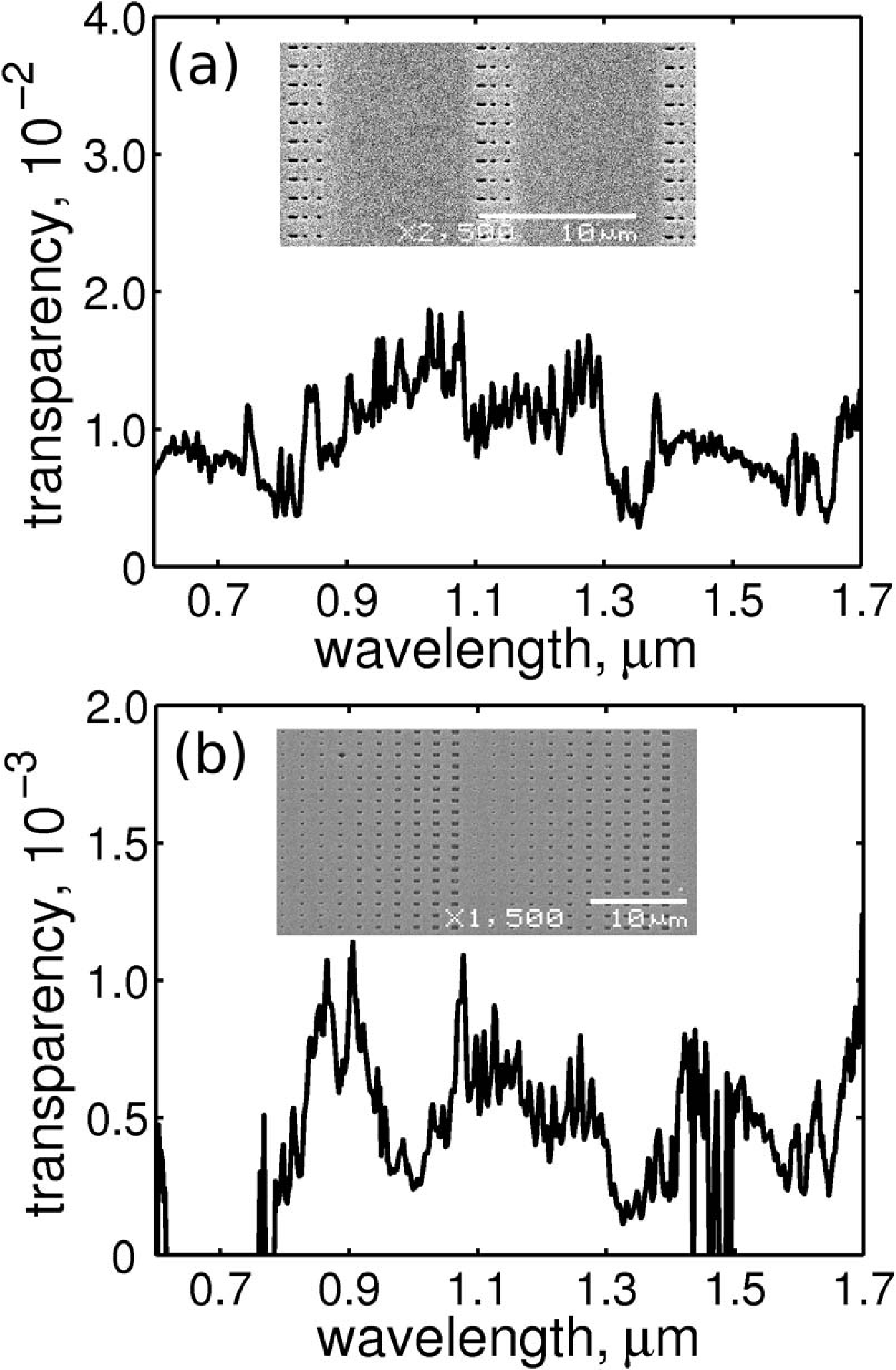}{quasi-periodic}{Spectra of the Thue-Morse pattern (a) and size-chirped (b) nanoholes lattices. Insets~-- scanning electronic microscope images of the corresponded structures.}

The measured transmission spectra of these two structures are shown in Fig.~\rpict{quasi-periodic}). For the quasi-periodic Thue-Morse structure, the transmission spectrum does not show any sharp peaks of transmission, indicating negligible surface plasmon excitation in the monitored wavelength range. In the chirped structure, the transmission spectrum is also strongly modified, however some weak peaks can be observed between 900\,nm and 1.1\,$\mu$m. This measurement indicates that chirping of the hole size in the lattice of square apertures on a metal film provides a possible way to tailor the transmission spectrum through metal perforated films.

\section{Conclusions}
\label{sec:Conclusions}



We have studied the transmission spectra of square lattices of nanoholes perforated in a gold film of thickness of 200\,nm. In our experiments with lattices of holes of 800\,nm size and period of $2\,\mu$m, we have measured transmission around 1\% and a flat spectrum in the region 600--1700\,nm. In contrast, in the experiments with the array of 300\,nm holes and the same period, we have observed a number of resonant transmission peaks at the wavelength ranges of 600--1000\,nm and 1.4--1.7\,$\mu$m. These transmission peaks are associated with the excitation of surface plasmons at the metal-air interface and lead to more than six times enhancement of transmission in comparison to the light transmitted through a single aperture of the same size. Our experimental results are found to be in a good qualitative agreement with the performed numerical calculations.

Finally, we demonstrate novel way to engineer the transmission through perforated metal films by inducing quasi-periodicity or chirping of the size of individual apertures. Such engineering may appear attractive for spectral filtering in spectroscopic and astronomical applications.

\section*{Acknowledgements}
The authors would like to thank Dr. Vince Craig for the help with AFM images. This research is supported by the Australian Research Council. The Australian National Fabrication Facility and AMMRF, funded through Australian Government's National Cooperation Research Infrastructure Strategy, is gratefully acknowledged for access to the facilities.



\end{sloppy}
\end{document}